\documentclass[useAMS,usenatbib]{mn2e}
\usepackage{graphics,graphicx}
\usepackage{aas_macros}
\usepackage{amssymb}
\usepackage{amsmath} 
\usepackage{latexsym}
\usepackage{color}
\usepackage{epstopdf}
\usepackage{marvosym}

\begin{document}

\title[The effect of CO$-$H$_2$O collisions in the rotational
  excitation of cometary CO]{The effect of CO$-$H$_2$O collisions in
  the rotational excitation of cometary CO}

\author[A. Faure, F. Lique and
  J. Loreau]{A. Faure$^{1}$\thanks{E-mail:
    alexandre.faure@univ-grenoble-alpes.fr},
  F. Lique$^{2}$\thanks{E-mail: francois.lique@univ-lehavre.fr}, and
  J.  Loreau$^{5}$\thanks{E-mail: jerome.loreau@kuleuven.be}, \\ $^1$
  Univ. Grenoble Alpes, CNRS, IPAG, F-38000 Grenoble, France \\ $^2$
  Normandie Universit\'e, Universit\'e du Havre and CNRS, LOMC,
  F-76063 Le Havre, France \\ $^3$ KU Leuven, Department of Chemistry, 3001 Heverlee, Belgium}

\date{Accepted ? Received ?}

\pagerange{\pageref{firstpage}--\pageref{lastpage}} \pubyear{2016}

\maketitle

\label{firstpage}

\begin{abstract}

We present the first accurate rate coefficients for the rotational
excitation of CO by H$_2$O in the kinetic temperature range
5-100~K. The statistical adiabatic channel method (SACM) is combined
with a high-level rigid-rotor CO$-$H$_2$O intermolecular potential
energy surface. Transitions among the first 11 rotational levels of CO
and the first 8 rotational levels of both {\it para}-H$_2$O and {\it
  ortho}-H$_2$O are considered. Our rate coefficients are compared to
previous data from the literature and they are also incorporated in a
simple non-LTE model of cometary coma including collision-induced
transitions, solar radiative pumping and radiative decay. We find that
the uncertainties in the collision data have significant influence on
the CO population distribution for H$_2$O densities in the range
$10^3-10^8$~cm$^{-3}$. We also show that the rotational distribution
of H$_2$O plays an important role in CO excitation (owing to
correlated energy transfer in both CO and H$_2$O), while the impact of
the ortho-to-para ratio of H$_2$O is found to be negligible.

\end{abstract}


\section{Introduction}

Rate coefficients (or cross sections) for molecular collisional
excitation are needed to model both thermal balance and spectral line
formation in astrophysical objects such as interstellar clouds,
protoplanetary disks and planetary/cometary atmospheres. In such
environments, the density can be so low ($n\ll 10^{10}$~cm$^{-3}$)
that collisions indeed cannot maintain a local thermodynamical
equilibrium (LTE) everywhere. Until recently, the accuracy of
theoretical collision data was mainly inferred indirectly by
comparison with related processes such as line broadening cross
sections \citep[see e.g][]{mengel00,wiesenfeld10}. In the last decade,
there has been huge progress in measuring state-to-state collision
cross sections at very low energy ($E_{col}\sim$1~meV) \citep[see][for
  a recent review]{naulin18}. Detailed comparisons between quantum
calculations and merged-beam measurements have thus allowed
experimentalists to verify theoretical cross sections in the 'cold'
regime (equivalent temperature $\sim$ 10~K) where quantum effects such
as resonances are predominant. Theory has passed all experimental
tests so far, establishing the reliability of modern collisional data
used in astrophysical models
\citep{chefdeville13,chefdeville15,gao19,bergeat19}. These theoretical
values have been obtained for 'simple' collision systems such as rigid
rotors excited by light atomic or diatomic perturbers, in particular
He atoms and H$_2$ molecules which are the most abundant collisional
partners in the interstellar medium (ISM).

However, collision rate coefficients are also needed for systems
involving heavy and even polyatomic perturbers, which challenge
current computational capabilities. For instance, in the thermosphere
of Titan, N$_2$ and CH$_4$ are the most abundant collisional
perturbers \citep{rezac13}. In cometary comas, collisional excitation
is generally dominated by H$_2$O (and/or electrons) and, for comets at
large heliocentric distances, by CO \citep{bockelee04}. The main
difficulty with such systems is the excessively large number of
angular couplings between the target and the projectile which make
full quantum calculations prohibitively expensive, both in terms of
computer time and memory requirements. As a result, to the best of our
knowledge, the only sets of rotationally-resolved rate coefficients
for the water perturber are those of \cite{green93} for CO$-$H$_2$O,
those of \cite{buffa00} for H$_2$O$-$H$_2$O and those of
\cite{dubernet19} for HCN$-$H$_2$O. The two former sets were estimated
using crude treatments while the latter was obtained using only
partially converged coupled-states (CS) calculations.

In this paper, we present the first quantum calculations for the
rotational excitation of CO by H$_2$O molecules. This is also the
first study to consider the two nuclear-spin isomers of H$_2$O, {\it
  ortho-} and {\it para-}H$_2$O (hereafter referred as o-H$_2$O and
p-H$_2$O), as two distinct perturbers. The statistical adiabatic
channel model (SACM) is used as a validated alternative to prohibitive
close-coupling calculations \citep{loreau18,loreau18b} and it is
combined here with the recent {\it ab initio} CO$-$H$_2$O interaction
potential computed by \cite{kalugina18}. The CO$-$H$_2$O collision
system is of great importance in comets but also in protoplanetary and
debris disks when the gas originates from ices in planetesimals
\citep{matra15}. In addition to providing collision data for the first
11 rotational levels of CO, we also present a non-LTE model for the CO
excitation in comets, including collision-induced transitions due to
p-H$_2$O and o-H$_2$O, radiative pumping of the $v=0\to 1$ band due to
solar photons and radiative decay due to spontaneous emission. A key
objective is to assess the impact of accurate CO$-$H$_2$O collisional
data on the rotational distribution of cometary CO. The next Section
describes the CO$-$H$_2$O scattering calculations. Section~3 compares
the present results with previous collisional data for CO. In
Section~4, we present simple nonequilibrium calculations for the
excitation of CO in comets. Concluding remarks are made in Section~5.

\section{Quantum statistical calculations}

The computation of inelastic cross sections from first principles
consists of two main steps. First, the electronic Schr\"odinger
equation is solved using quantum chemistry methods. Within the
Born-Oppenheimer approximation, a single intermolecular potential
energy surface (PES) governing the motion of the nuclei is defined as
function of nuclear coordinates. The nuclear motion is solved
separately, in a second step, using quantum or (semi)classical
scattering methods. We describe in this section our calculations for
the CO$-$H$_2$O system.

\subsection{Potential energy surface}

The CO$-$H$_2$O scattering calculations presented below are based on
the high-level {\it ab initio} intermolecular potential recently
determined by \cite{kalugina18}. The interacting molecules CO and
H$_2$O are considered as rigid-rotors, with geometries corresponding
to their ground vibrational state. The PES was calculated at 93,000
relative orientations using the explicitly-correlated single- and
double-excitation coupled cluster theory with a non-iterative
perturbative treatment of triple excitations [CCSD(T)-F12a] along with
the augmented correlation-consistent aug-cc-pVTZ basis sets (AVTZ
hereafter). The intermolecular energies were then fitted to a
body-fixed expansion expressed as a sum of spherical harmonic products
and a Monte-Carlo estimator was employed to select the largest
expansion terms. This CO$-$H$_2$O PES has a global well depth
$D_e=646.1$~cm$^{-1}$ for an intermolecular center-of-mass separation
$R=7.42$~$a_0$ and a planar configuration OC$-$H$_2$O, with a weak
hydrogen bond between the C atom of CO and one H atom of
H$_2$O. Bound-state rovibrational energy calculations using this PES
have been found in very good agreement with experiments
\citep{barclay19}, confirming the good accuracy of the potential.  We
note that a full-dimensional CO$-$H$_2$O PES was recently computed at
the same level (CCSD(T)-F12a/AVTZ) by \cite{liu19}. These authors
found a similar but deeper well depth ($D_e=685.5$~cm$^{-1}$), which
is due to the basis set superposition error (BSSE) not corrected by
\cite{liu19}.

\subsection{The statistical adiabatic channel model}

Rigorous close-coupling quantum scattering calculations become
impracticable when the number of coupled channels exceed typically
10,000. In such situation, statistical methods may provide a useful
alternative. Statistical theories generally apply when a collisional
system can access a deep well on the PES and the lifetime of the
complex is sufficiently long. We have recently \citep{loreau18}
developed a statistical approach inspired by the statistical adiabatic
channel model (SACM) of \cite{quack75}. In our implementation, we
diagonalize the Hamiltonian excluding the kinetic term in the basis of
rotational functions for the collision. In the present case, the
angular functions are products of Wigner rotation matrices for H$_2$O
and spherical harmonics for CO as well as for the relative orbital
motion \citep{phillips95}. This expansion corresponds to the coupling
of the angular momenta ${\boldsymbol j_1}$ of H$_2$O, ${\boldsymbol
  j_2}$ of CO, and ${\boldsymbol \ell}$ for the relative motion to
form the total angular momentum $\boldsymbol{ J = j_1+j_2+\ell}$. Upon
diagonalization, for each value of $J$ we obtain a set of adiabatic
rotational curves as a function of the intermolecular distance $R$
with an asymptotic value given by the sum of rotational energies of
H$_2$O and CO. These curves possess centrifugal barriers that depend
on the orbital quantum number $\ell$, and for a given initial state,
the collision is assumed to take place if the collision energy is
larger than the height of the barrier. The probability is then divided
equally among the channels for which the energy is larger than their
respective centrifugal barriers, allowing us to compute state-to-state
cross sections and rate coefficients.

This procedure was found to provide rate coefficients with an average
accuracy better than 50\% for ion-molecule systems up to room
temperature \citep{loreau18}. In the case of CO$-$H$_2$O, the SACM
{\it partial} rate coefficients (computed for a total angular momentum
$J=0$) were shown to be accurate to within 40\% up to 100~K
\citep{loreau18b}. Full comparisons between SACM and close-coupling
calculations can be found in the two papers by Loreau and co-workers.

In this paper the cross sections were calculated for transitions among
the first 11 energy levels of CO (up to $j_2=10$) and the first 8
energy levels of both p-H$_2$O and o-H$_2$O (up to $j_1=3$) and for
collision energies between 0 and 500~cm$^{-1}$ in order to derive rate
coefficients up to $T=100$ K. This requires to obtain all the
adiabatic curves up to a total energy of 750~cm$^{-1}$. The basis set
of angular functions included levels up to $j_1=5$ and $j_2=15$, and
tests were performed to ensure the convergence of the adiabatic curves
with respect to the size of the basis set. The rotational constant of
CO was taken to be $B_0=1.92253$ cm$^{-1}$, while the energy of the
rotational levels of H$_2$O, $E_{j_1k_ak_c}$, was obtained using the
effective Hamiltonian of \cite{kyro81}. They are labeled with $j_1$,
$k_a$, and $k_c$, where $k_a$ and $k_c$ are pseudoquantum numbers
corresponding to the projection of the angular momentum $j_1$ along
the axis of least and greatest moment of inertia,
respectively. Adiabatic curves were generated for values of the total
angular momentum from $J=0$ up to $J=100$ for o-H$_2$O and p-H$_2$O
separately. High total angular momentum values were found to be
necessary as the cross section converges very slowly with $J$ due to
the deep well in the PES as well as to the large reduced mass of
H$_2$O-CO.


Two sets of rate coefficients describing the excitation of CO by
H$_2$O were considered: the first consists of rate coefficients for
which H$_2$O is in the ground rotational state of its nuclear-spin
isomer, i.e. $k_{j_2\rightarrow j_2^\prime}(T)=k_{0_{00}, j_2
  \rightarrow 0_{00}, j_2^\prime}(T)$ or $k_{1_{01}, j_2 \rightarrow
  1_{01}, j_2^\prime}(T)$ for p-H$_2$O and o-H$_2$O, respectively.
This set of ``ground-state'' rate coefficients neglects the
simultaneous excitation of CO and H$_2$O as well as the fact that
excited rotational states of H$_2$O can be significantly
populated. Such an approximation is therefore expected to be accurate
only at very low temperatures.

To take into account the transitions in H$_2$O, we constructed a
second set of rate coefficients (separately for p-H$_2$O and o-H$_2$O)
by summing over all possible final states of H$_2$O, and averaging
over the initial rotational states of H$_2$O assuming a thermal
population distribution:
\begin{equation}\label{thermalized_rates}
k_{j_2\rightarrow j_2^\prime}(T)
= \sum_{j_1k_ak_c j_1^\prime k_a^\prime k_c^\prime} P_{j_1k_ak_c}(T) k_{j_1k_ak_c, j_2\rightarrow j_1^\prime k_a^\prime k_c^\prime,  j_2^\prime}(T),
\end{equation}
where
\begin{equation}
  P_{j_1k_ak_c}(T)=\frac{g_{j_1}\exp(-E_{j_1k_ak_c}/k_BT)}{\sum_{j_1k_ak_c}
    g_{j_1}\exp(-E_{j_1k_ak_c}/k_BT)}
\end{equation}
is the Boltzmann population of level $j_1k_ak_c$ at temperature $T$
and $k_B$ is the Boltzmann constant. This set of ``thermalized'' rate
coefficients assumes that the kinetic and rotational temperatures are
equal, and satisfies detailed balance. These thermalized rate
coefficients can be much larger than the ground-states values, even at
low temperature, owing to correlated rotational energy transfer in
both collision partners (see below). All de-excitation rate
coefficients are available as supplementary material. Excitation rate
coefficients can be derived using the detailed balance principle.

\section{Results}

\begin{figure}
\includegraphics*[width=8.5cm,angle=-90.]{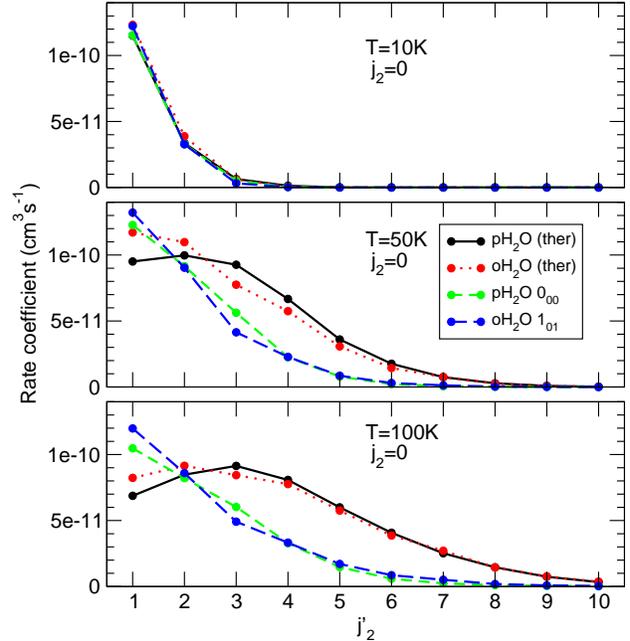}
\caption{SACM rate coefficients for the rotational excitation of CO by
  H$_2$O from the ground state ($j_2$=0) to the lowest ten excited
  states. The projectiles p-H$_2$O and o-H$_2$O are assumed to be either
  thermalized at the kinetic temperature (10, 50 or 100~K) or in their
  ground rotational states $0_{00}$ and $1_{01}$, respectively. See
  text for details.}
\label{fig1}
\end{figure}

Rate coefficients for the rotational excitation of CO by H$_2$O are
presented below at three kinetic temperatures, 10, 50 and 100~K, for
transitions out of the ground rotational state of CO ($j_{2}$=0) as
functions of the final $j_2$ ($j^\prime_2=1-10$). In Fig.~\ref{fig1},
four sets of data are presented, corresponding to the ``ground-state''
and ``thermalized'' SACM rate coefficients defined above with p-H$_2$O
and o-H$_2$O as colliders. It is seen that all four sets have very
similar rate coefficients at 10~K and that the transition $j_2=0\to 1$
dominates. Small differences between thermalized and ground-state
rate coefficients were expected at 10~K because {\it i)} only the
ground-state of p-H$_2$O and o-H$_2$O is significantly populated at
thermal equilibrium and {\it ii)} the excitation of H$_2$O is
inhibited because the rotational thresholds are larger than 10~K (23~K
and 54~K for the first excitation in o-H$_2$O and p-H$_2$O,
respectively). In fact, however, substantial differences (more than a
factor of 2) are found for transitions with $j^\prime_2>3$ whose rate
coefficients are smaller than $10^{-11}$~cm$^3$s$^{-1}$. For instance,
the thermalized and ground-state rate coefficients for the excitation
$j_2=0\to 4$ ($\Delta E$=38.4~cm$^{-1}$) by p-H$_2$O are $1.26\times
10^{-12}$~cm$^3$s$^{-1}$ and $3.16\times 10^{-13}$~cm$^3$s$^{-1}$,
respectively. The much larger thermalized value is due to the
correlation between the CO excitation $j_2=0\to 4$ ($\Delta
E$=38.4~cm$^{-1}$) and the simultaneous H$_2$O de-excitation
$j_1k_ak_c=1_{11}\to 0_{00}$ ($\Delta E$=37.1~cm$^{-1}$). This
particular process provides a major contribution to the thermalized
rate coefficient, despite the low population of the $1_{11}$ level at
10~K ($\sim$1\%). We also note that the near-resonant character of
this correlated (de)excitation is not accounted for by SACM
calculations and that close-coupling calculations would predict an
even larger thermalized rate coefficient. At 50 and 100~K, the
differences between the thermalized and the ground-state rate
coefficients are clearly visible, reflecting both the larger
population of excited H$_2$O levels (thus pair correlations) and the
opening of excitation channels in H$_2$O. In particular, the
transitions $j_2=0\to 2, 3$ are preferred in the thermalized data at
100~K while $j_2=0\to 1$ is favored in the ground-state data. We also
note that the difference between p-H$_2$O and o-H$_2$O is very small,
both for thermalized and ground-state data, suggesting that the
ortho-to-para ratio (OPR) of H$_2$O should not play a major role in
the excitation of CO (see Section~4 below).

\begin{figure}
\includegraphics*[width=8.5cm,angle=-90.]{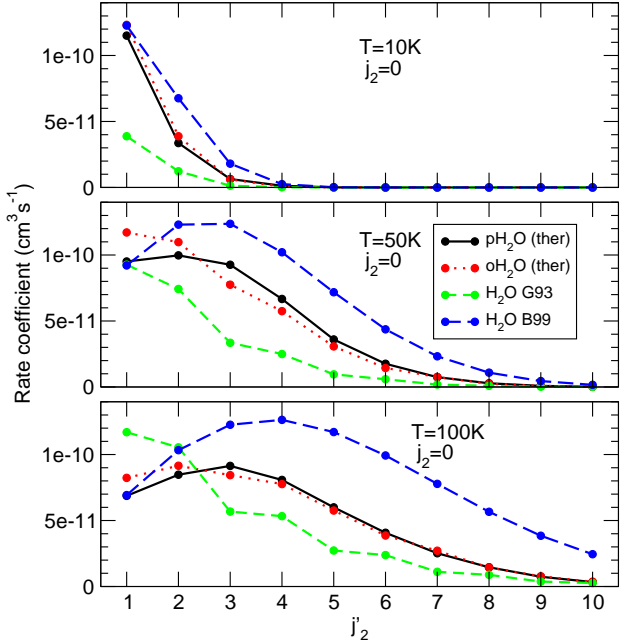}
\caption{Same as Fig.~\ref{fig1}. The thermalized SACM data for
  p-H$_2$O and o-H$_2$O are compared to the data of Biver et al. (1999)
  and Green (1993). See text for details.}
\label{fig2}
\end{figure}

In Fig.~\ref{fig2}, the above set of thermalized rate coefficients is
compared to previous data from the literature: those of \cite{biver99}
(denoted as B99) and those of \cite{green93} (denoted as
G93). \cite{biver99} assumed a {\it total} cross-section of
$\sigma_c=2\times 10^{-14}$~cm$^{-2}$, with no energy dependence. This
value was derived from room temperature pressure broadening
measurements on H$_2$O-H$_2$O \citep{biver97}, with an estimated
uncertainty of a factor of $\sim 2$. The total collisional rate
coefficient was computed as $k_{c}(T)=\sigma_c\langle v(T) \rangle$,
where $T$ is the kinetic temperature and $\langle v(T) \rangle$ is the
thermal mean velocity of the CO molecule relative to the cometary
gas. State-to-state rate coefficients from an initial CO level $j_2$
to a final level $j^\prime_2$ were then obtained as $k_{j_2\to
  j^\prime_2}(T) = k_cP_{j^\prime_2}(T)$, where $P_{j^\prime_2}(T)$ is
the Boltzmann population of the final level $j^\prime_2$ at
temperature $T$. This recipe corresponds to assuming that each
collision redistributes the molecule according to the Boltzmann
distribution, which has no physical basis, except that it satisfies
the detailed balance principle (by construction). \cite{green93}, on
his side, combined line shape measurements on N$_2$-H$_2$O with a
rate-law derived from the infinite-order-sudden (IOS)
approximation. The total excitation rates were expected to be rather
accurate (about 10\%) at room temperature, where the rate-law
parameters were fitted. It should be noted that in contrast to our
SACM calculations, the B99 and G93 collision data sets do not
distinguish between p-H$_2$O and o-H$_2$O. It is found in
Fig.~\ref{fig2} that the magnitude and the $\Delta j_2=j^\prime_2-j_2$
dependence of the rate coefficients are reasonably described by the
B99 and G93 data sets: the largest rate coefficients are $\sim
10^{-10}$~cm$^3$s$^{-1}$ and they (globally) slowly decrease with
increasing $\Delta j_2$. The overall agreement is somewhat surprising,
in particular for the B99 data set given the simplicity of the
prescription. We note, however, that differences larger than a factor
of 2 are observed for transitions with $\Delta j_2>2$ and, more
generally, that the B99 set overestimates while the G93 set tends to
underestimate the SACM rate coefficients.
 
\begin{figure}
\includegraphics*[width=8.5cm,angle=-90.]{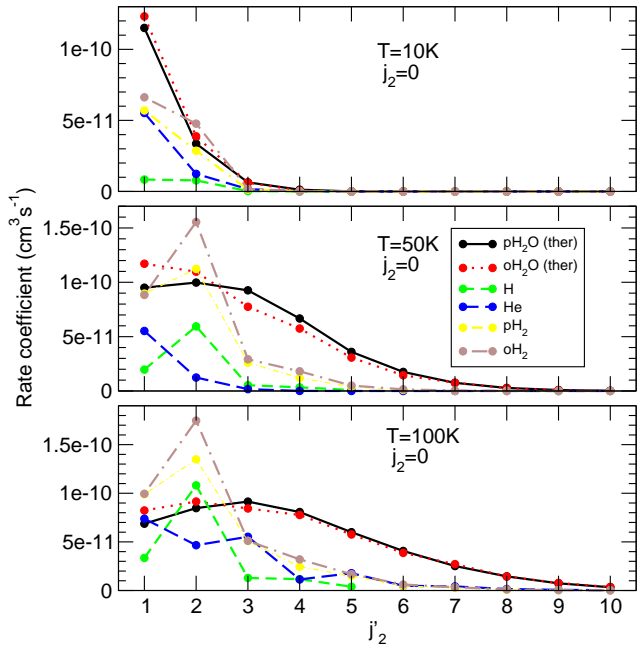}
\caption{Same as Fig.~\ref{fig1}. The thermalized SACM data for
  p-H$_2$O and o-H$_2$O are compared to the data for other projectiles:
  H, He, p-H$_2$ and o-H$_2$.}
\label{fig3}
\end{figure}

It is now interesting to compare the efficiency of H$_2$O relative to
other colliders for exciting CO. In Fig.~\ref{fig3}, the thermalized
SACM rate coefficients for CO$-$H$_2$O are compared to the
corresponding data for CO$-$H \citep{yang13}, CO$-$He
\citep{cecchi02}, CO$-$p-H$_2$ and CO$-$o-H$_2$ \citep{yang10}. These
latter four sets of data were obtained from quantum close-coupling
calculations combined with modern PES. Their accuracy at the
state-to-state level is expected to be very good (about 20\%), as
verified experimentally for CO$-$He and CO$-$H$_2$
\citep{carty04,chefdeville15}. We can first observe that, at 10~K,
p-H$_2$O and o-H$_2$O are the most effective colliders with a rate
coefficient for the transition $j_2=0\to 1$ that is a factor of $\sim
2$ larger than those for the other colliders. Atomic hydrogen is the
less effective, with rate coefficients lower than
$10^{-11}$~cm$^3$s$^{-1}$. At larger temperatures, the $\Delta j_2$
dependence of the rate coefficients is found to be very different for
H$_2$O than for H and H$_2$: while the water rate coefficients slowly
decrease with increasing $\Delta j_2$, a strong propensity rule with
$\Delta j_2=2$ is observed for atomic and molecular hydrogen. As a
result, the largest rate coefficients at 100~K are those of H and
H$_2$ for the transition $j_2=0\to 2$. For transitions with $\Delta
j_2>2$, however, H$_2$O is seen to be more than twice as effective as
the other colliders. The magnitude of the rate coefficients for the
excitation of CO by H$_2$O is thus rather similar to that of other
neutral colliders. The main difference is the lack of strong
propensity rules and the relatively slow decrease of the rate
coefficients with increasing $\Delta j_2$. This reflects mainly the
larger potential well depth but also the fact that SACM calculations
are not able to catch resonance and interference effects (Loreau et
al. 2018a, 2018b). In particular, as discussed above, SACM
calculations are expected to underestimate near-resonant rotational
energy transfers.

We note, finally, that free electrons provide another source of CO
rotational excitation in astronomical environments. Owing to the small
dipole of CO (0.11~D), however, rate coefficients for electron-impact
rotational excitation do no exceed $\sim 2\times
10^{-8}$~cm$^3$s$^{-1}$. This means that electrons compete with
neutrals in exciting CO only when the electron-to-neutral number
density ratio is larger than $\sim 10^{-2}$.

\section{Non-LTE model of CO in comets}

In this Section, we present a simple non-LTE model of CO excitation in
a cometary coma. Our main objective is to assess the impact of the
collisional rate coefficients on the predicted rotational populations
of CO. We have chosen physical conditions corresponding to an {\it
  average} comet located at $R_{\rm h}$=1~AU from the Sun with a total
production rate from the nucleus $Q_{\rm H_2O}=1\times
10^{29}$~s$^{-1}$, an expansion velocity $v_{\rm exp}=0.8$~km~s$^{-1}$
and a neutral gas temperature $T_k$ in the range $10-100$~K. Water is
assumed to be the dominant neutral collider in the coma. The H$_2$O
density can be described by a standard isotropic Haser model
\citep{haser57}:
\begin{equation}
n_{\rm H_2O}(r)=\frac{Q_{\rm H_2O}}{4\pi r^2 v_{\rm
    exp}}\exp\left(-r\frac{\beta_{\rm H_2O}}{v_{\rm exp}}\right),
\label{Haser}\end{equation}
where $r$ is is the nucleocentric distance and $\beta_{\rm H_2O}$ is
the H$_2$O photodissociation rate at $R_{\rm h}$=1~AU. The coupled
radiative transfer and statistical equilibrium (SE) equations are
solved with the public version of the \texttt{RADEX} code using the
large velocity gradient (LVG) approximation. The problem is treated as
both local and steady-state by running a grid of \texttt{RADEX}
calculations covering a range of H$_2$O densities. We thus assume that
the physical processes driving the population are much faster than
variations in physical conditions of the coma. Note that we also
neglect CO photodissociation (the CO lifetime is about $5.3\times
10^5$~s at $R_{\rm h}$=1~AU for the active Sun \citep{huebner92}) so
that chemical processes (formation and destruction of CO) are omitted
from the SE equations. This approach may not be valid over the whole
coma and more elaborate and accurate non-LTE treatments exist in the
literature \citep{zakharov07,yamada18}. We believe, however, that the
main features of the collisional effects are well captured by our
simple model.

\begin{figure}
\includegraphics*[width=8.5cm,angle=-90.]{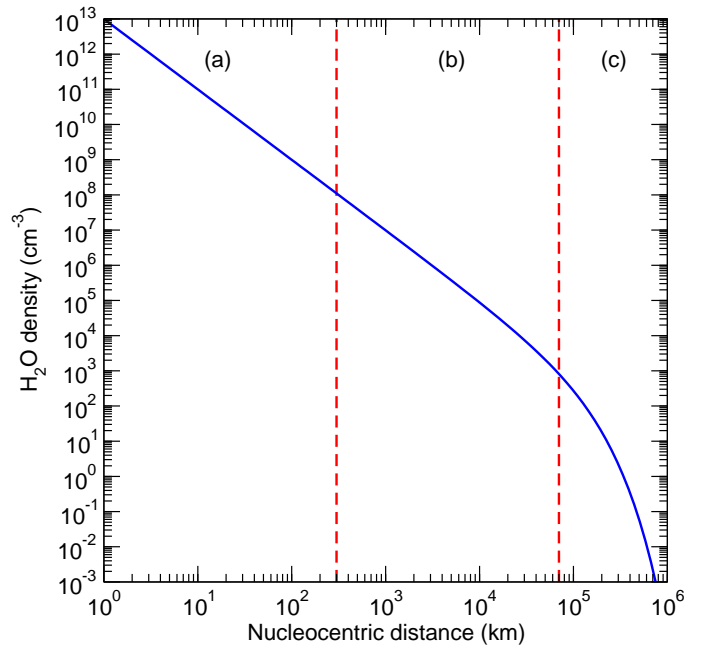}
\caption{Coma density profile of H$_2$O in the Haser's model. The
  photodissociation rate was taken as $\beta_{\rm H_2O}=1.042\times
  10^{-5}$~s$^{-1}$ from Zakharov et al. (2007). The dashed lines
  delineate, for a production rate $Q_{\rm H_2O}=1\times
  10^{29}$~s$^{-1}$, the three approximate density regimes: (a) LTE
  equilibrium, (b) non-LTE regime, and (c) fluorescence
  equilibrium. See text for details.}
\label{fig4}
\end{figure}

The input parameters to \texttt{RADEX} are the kinetic temperature,
$T_k$, the column density of CO, $N$(CO), the line width, $\Delta v$,
and the density of the collider, here $n_{\rm H_2O}$. Three kinetic
temperatures were selected: 10, 50 and 100~K. The column density was
fixed at $N({\rm CO})=10^{14}$~cm$^{-2}$, which is the typical
magnitude at $r\sim 1000$~km for a CO/H$_2$O abundance ratio of $\sim
10$\% \citep[e.g.][]{lupu07}. We have checked, however, that the
relative populations are not sensitive to the column density provided
that $N({\rm CO})< 10^{16}$~cm$^{-2}$ (i.e. the lines are optically
thin). The water density is varied in the large range
$10^{-3}-10^{13}$~cm$^{-3}$ corresponding to nucleocentric distances
between 1 and $10^6$~km in the Haser model, as plotted in
Fig.~\ref{fig4}. The line width is fixed at 1.6~km~s$^{-1}$,
i.e. $\Delta v=2\times v_{\rm exp}$. Finally, the background radiation
field includes both the 2.73~K cosmic microwave background (CMB) and
the Sun radiation. This latter contribution is taken as a diluted
blackbody of 5770~K with the dilution factor $W=\Omega_{S}/4\pi$,
where $\Omega_{S}\sim 6.79\times 10^{-5}$~sr is the Sun solid angle
seen at $R_{\rm h}$=1~AU.

Input data include the CO energy levels $(v_2, j_2)$ ($v_2$ is the CO
vibrational quantum number), the spontaneous emission Einstein
coefficients and the collisional rate coefficients. Level energies and
Einstein coefficients were taken from the \texttt{HITRAN}
database\footnote{https://hitran.org/.} \citep{gordon17}. Only the
first excited vibrational level $v_2=1$ was taken into account because
the solar excitation rate of $v_2=1$ is at least two orders of
magnitude greater than those of $v_2>1$ \citep{crovisier83}. Our model
thus include the lowest 80 ro-vibrational levels of CO, i.e. up to
level $(v_2, j_2)=(1, 32)$ which lies 4148~cm$^{-1}$ above the
ground-state $(0, 0)$. Since our SACM rate coefficients are available
for the lowest 11 rotational levels only, we had to extrapolate the
collisional data above $j_2=10$. It should be noted, however, that the
corresponding levels are negligibly populated (less than 1\%) at the
investigated temperatures and the extrapolated collisional data play
only a minor role in the SE equations. In practice, the {\it total}
de-excitation rotational rate coefficients were simply fixed at
$10^{-10}$~cm$^3$s$^{-1}$ with an equal distribution among final
levels and no temperature dependence. Similarly the {\it total}
de-excitation vibrational rate coefficients were fixed at
$10^{-13}$~cm$^3$s$^{-1}$ (with again an equal distribution among
final levels), as suggested by the CO($v_2=1$)$-$H$_2$O quenching
measurements of \cite{wang99} performed at 300~K. Finally, for pure
rotational transitions involving rotational levels $j_2\leq 10$ within
$v_2=1$, we adopted the corresponding rate coefficients within
$v_2=0$. Rotational rate coefficients indeed depend only weakly on the
vibrational state \citep{carty04}. This extrapolation procedure was
used for the SACM, the B99 and the G93 collision data sets.

\begin{figure}
\includegraphics*[width=8.5cm,angle=-90.]{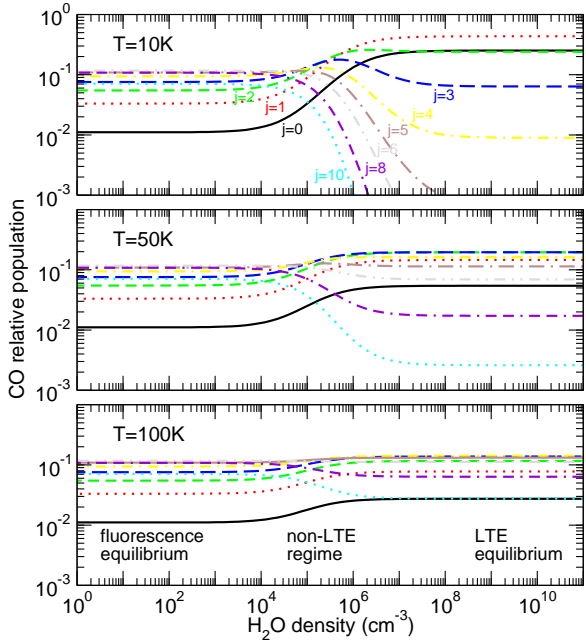}
\caption{Level populations of CO ($j=0-6, 8, 10$) as functions of
  H$_2$O density for kinetic temperatures of 10, 50 and 100~K. The
  thermalized SACM rate coefficients are employed with an OPR of
  H$_2$O fixed at 3.}
\label{fig5}
\end{figure}

The population distribution of the CO rotational levels $j_2$ in
$v_2=0$ is thus governed by the balance between the collision-induced
transitions, the vibrational excitation ($v_2=0 \to 1$) by the solar
infrared radiation field, the rotational excitation by the CMB
radiation field and the radiative decay due to spontaneous
emission. The results of our non-LTE model are presented in
Fig.~\ref{fig5} where the population of levels $j_2$ is plotted as
function of the H$_2$O density for the selected kinetic temperatures
10, 50 and 100~K. The set of thermalized SACM rate coefficient is
employed here with an OPR for H$_2$O equal to 3. We can clearly
observed the presence of three different regimes: the fluorescence
equilibrium at low densities ($n_{\rm H_2O}\lesssim 10^3$~cm$^{-3}$),
the non-LTE regime in the range $10^3\lesssim n_{\rm H_2O}\lesssim
10^8$~cm$^{-3}$ and the LTE-regime at larger densities. Thus, for a
production rate $Q_{\rm H_2O}=1\times 10^{29}$~s$^{-1}$, LTE applies
at $r\lesssim 300$~km, i.e. in the inner coma, while the fluorescence
equilibrium distribution is completely established at $r\gtrsim
7\times 10^4$~km (see Fig.~\ref{fig4}). The non-LTE regime extends
typically from 300 to 70,000~km, representing about 10\% of all CO
molecules in the coma. We can notice in Fig.~\ref{fig5} that the
non-LTE regime of densities does not significantly vary with
temperature. In contrast, the LTE population distribution is very
sensitive to the temperature: at 10~K significant populations ($>$1\%)
are found only for $j_2=0-3$ while at 100~K all levels $j_2=0-10$ are
occupied above 2\%.

\begin{figure}
\includegraphics*[width=8.5cm,angle=-90.]{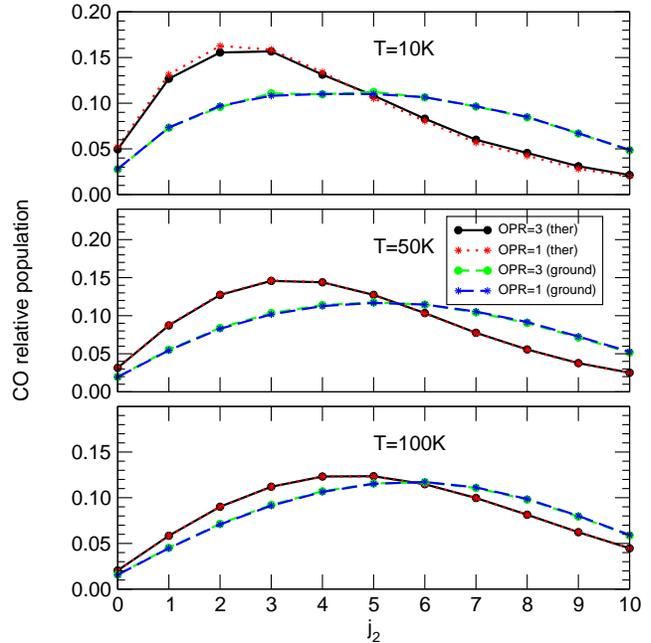}
\caption{CO population distribution at a fixed H$_2$O density of $1.9
  \times 10^5$~cm$^{-3}$ for the three kinetic temperatures 10, 50 and
  100~K. The SACM thermalized and ground-state rate coefficients are
  employed with an OPR of H$_2$O equal to 1 and 3. See text for
  details.}
\label{fig6}
\end{figure}

In the following, the H$_2$O density is fixed at $n_{\rm
  H_2O}=1.9\times 10^5$~cm$^{-3}$, which is representative of the
non-LTE regime. The corresponding nucleocentric distance is $r\sim
7000$~km. In Fig.~\ref{fig6} the CO relative population is plotted as
function of $j_2$ for the two sets of SACM rate coefficients
(thermalized and ground-state) and for an OPR of H$_2$O equal to 1 and
3. The OPR of H$_2$O in comets is debated but values are generally
comprised between 1 and 3, with a median equal to 2.9 \citep[see][and
  references therein]{faure19}. We first observe that the impact of
the OPR of water is negligible, as expected from the small differences
seen in Fig.~\ref{fig1}. It can be also noticed that the thermalized
SACM rate coefficients predict a more peaked distribution than the
ground-state rate coefficients, with a greater population for low
$j_2$ levels. The relative difference is found to reach $\sim 60$\% at
10~K for $j_2=2$. This effect is due to pair correlations between CO
and H$_2$O transitions, as discussed in Section~3. Perhaps
surprisingly, the influence of the H$_2$O rotational distribution is
found to be less important at higher temperature. In fact, this mainly
reflects the LTE distributions which are closer to the fluorescence
equilibrium distribution at 50~K and mostly 100~K than at 10~K,
attenuating the impact of collisional data.

\begin{figure}
\includegraphics*[width=8.5cm,angle=-90.]{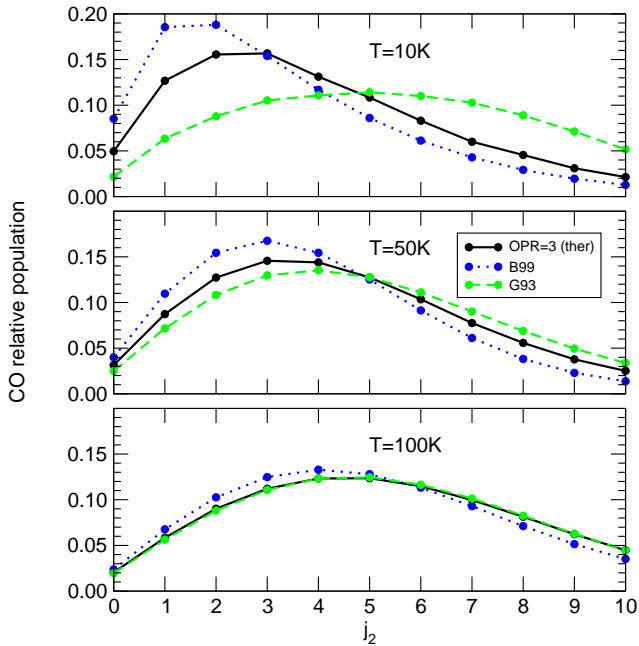}
\caption{Same as Fig.~\ref{fig6}. Three sets of collisional rate
  coefficients are employed: the thermalized SACM set with the OPR of
  H$_2$O fixed at 3, the B99 set and the G93 set. See text for
  details.}
\label{fig7}
\end{figure}

Finally, in Fig.~\ref{fig7} we compare the CO rotational populations
as predicted by our thermalized SACM rate coefficients (with an OPR of
H$_2$O fixed at 3) and by the B99 and G93 rate coefficients. As in
Fig.~\ref{fig6}, the influence of collisional data is found to
decrease with temperature, which again reflects the relative deviation
between the LTE and the fluorescence equilibrium distribution. At
10~K, the uncertainties in collisional data have the largest impact
with relative differences up to a factor of $\sim 3$ for $j_2=1$. The
B99 set overestimates while the G93 set underestimates the populations
of levels $j_2\lesssim 4$, which is consistent with the trend in the
rate coefficients shown in Fig.~\ref{fig2}. For levels with $j_2>4$,
however, the reverse is observed, emphasizing the role of collisional
transitions out of $j_2>0$.

In summary, using a simple non-LTE model of CO in standard cometary
conditions, we have found that uncertainties in the collisional rate
coefficients have a significant impact on the CO population
distribution in the non-LTE regime ($n_{\rm H_2O}\sim
10^5$~cm$^{-3}$). In particular, the CO rotational populations
predicted by three different collisional data sets were found to
differ by up to a factor of 3 at 10~K. Thanks to our SACM rate
coefficients, we were also able to demonstrate that the H$_2$O
rotational distribution plays an important role, even at low
temperature (due to correlated rotational energy transfers) while the
OPR of H$_2$O has an almost negligible influence.

\section{Concluding remarks}

We have provided the first quantum rate coefficients for the
rotational excitation of CO by p-H$_2$O and o-H$_2$O. These rate
coefficients were obtained by combining SACM calculations with a
recent {\it ab initio} CO$-$H$_2$O PES computed at the
CCSD(T)-F12a/AVTZ level. Rotational transitions among the first 11
energy levels of CO and the first 8 energy levels of both p-H$_2$O and
o-H$_2$O were considered. Two different sets of rate coefficients were
obtained: one corresponding to p-H$_2$O and o-H$_2$O in their ground
rotational state and one corresponding to thermal distributions of
each nuclear-spin isomer. Correlated energy transfer in both CO and
H$_2$O was shown to be important while the differences between
p-H$_2$O and o-H$_2$O were found to be small. Our set of
``thermalized'' rate coefficients was compared to previous data from
the literature and significant differences were observed. We also
found that H$_2$O is more effective than other colliders (H, He and
H$_2$) for exciting CO transitions with $\Delta j_2>2$. Finally, we
have assessed the impact of the collision data on the CO population
distribution in a cometary coma using a simple non-LTE model including
collision-induced transitions, radiative pumping and radiative
decay. We have found that the uncertainties in the collision data can
affect the rotational populations by large factors at low temperature
and at H$_2$O densities $\sim 10^3 - 10^8$~cm$^{-3}$. In addition, the
rotational distribution of H$_2$O was found to play a significant role
so that a consistent non-LTE model of CO excitation should include a
realistic (i.e. non-LTE) population distribution for H$_2$O. This in
turn would require the knowledge of accurate H$_2$O-H$_2$O rotational
rate coefficients, which are lacking in the literature. On the other
hand, the OPR of H$_2$O was shown to have a negligible impact on the
CO excitation, which removes one uncertainty from the interpretation
of CO spectra.

Future studies will need to consider other linear targets such as CN,
CS, HCO$^+$ as well as non-linear species such as NH$_3$, H$_2$CO,
etc. colliding with both H$_2$O and CO. The present SACM approach
should be therefore extended to treat two non-linear polyatomic
collision partners. We note that many data are available for the
electron-impact rotational excitation of cometary molecules
\cite[e.g.][]{harrison13}. Comparing the efficiency of electrons
relative to neutrals (H$_2$O and CO) will be very useful to
distinguish between the different collisional non-LTE regimes in a
comet (H$_2$O-, $e$- or CO-dominated). Finally, we strongly encourage
modellers to use the present CO$-$H$_2$O data in any model of CO
excitation in comets and, more generally, to use the ever-increasing
amount of theoretical collision data available for astrophysical
applications.

\section*{Acknowledgements}

This research was supported by the CNRS Programme National `Physique
et Chimie du Milieu Interstellaire' (PCMI) of CNRS/INSU with INC/INP
co-funded by CEA and CNES . Nicolas Biver, J\'er\'emie Boissier and
Lucas Paganini are acknowledged for very useful discussions.

\bibliographystyle{mn2e}

\bibliography{faure1012}

\bsp

\label{lastpage}

\end{document}